**Self-organized Models of Selectivity in Ca and Na Channels**

Talk for IMA Workshop on Solvation†

Bob Eisenberg§

November 19, 2008

Selectivity is one of the most important properties of living systems. One of the founders of molecular biology (Nobel Laureate Aaron Klug) recently said (with some hyperbole I suspect [1]) "There is only one word that matters in biology and that is specificity."

My collaborators and I [2] study selectivity in ion channels. Ion channels are proteins with a hole down their middle that are the (nano nearly pico)valves of life. Ion channels control an enormous range of biological function in health and disease. A large amount of data is available about selectivity in many channels. Selectivity in ion channels occurs without structural change of the channel protein (on the biological time scale of $10^{-5}$ sec or longer) and does not involve changes in covalent bonds (i.e., changes in shape of electron orbitals). Selectivity in channels involves only electrodiffusion—usually of charged hard spheres. Thus, physical analysis of selectivity in ion channels is easier than analysis of specificity in enzymes or many other proteins while being at least as important biologically.

A simple pillbox model with two adjustable parameters accounts for the selectivity of both DEEA (Aspartate Glutamate Glutamate Alanine) Ca channels [3,4] and DEKA (Aspartate Glutamate Lysine Alanine) Na channels [5] in many ionic solutions of different composition and concentration. The predicted properties of the Na and Ca channels are very different even though 'Pauling' crystal radii are used for ions and all parameters are the same for both channels in all solutions. Only the side chains are different in the model of the Ca and Na channels. No information from crystal structures is used in the model. Side chains of the channel protein are grossly approximated as solid spheres.

How can such a simple model give such powerful results when chemical intuition says that selectivity depends on the precise relation of ions and side chains? We [1] use Monte Carlo simulations of this model that determine the most stable-lowest free energy-structure of the ions and side chains. **Structure is the computed consequence of the forces in this model and so is different in different conditions.** The relationship of ions and side chains vary with ionic solution and are very different in simulations of the Na and Ca channels. Selectivity is a consequence of the 'induced fit' of side chains to ions and depends on the flexibility (entropy) of the side chains as well as their location. The induced fit depends on the concentrations of ions in the surrounding solutions in a complex way. Thus, calculations in a single set of conditions are of limited use. In particular, calculations of 'free energy of binding' in infinitely dilute or ideal solutions are not likely to give useful estimates of binding in physiological solutions. Physiological solutions are typically ~ 200 mM, far from dilute.

The self-organized induced-fit model captures the relation of side chains and ions well enough to account for selectivity of both Na channels and Ca channels in the wide range of conditions measured in experiments, even though the components of the model are grossly oversimplified. Perhaps the simplified model works because the structures in both the model and the real channel are the most stable, self-organized, and at their free energy minimum, different in different conditions.

It seems that an important biological function can be understood by an oversimplified model if the model calculates the 'most stable' structure as it changes from solution to solution, and mutation to mutation.

†Available online from Institute for Mathematics and its Applications (IMA) University of Minnesota http://www.ima.umn.edu/ at Eisenberg Talk IMA November 19 2009

§ beisenbe@rush.edu, Department of Molecular Biophysics and Physiology, Rush University Medical Center, Chicago IL 60612.